\documentclass[english,aps,prb,preprint,superscriptaddress,showpacs,showkeys]{article}
\usepackage[T1]{fontenc}
\usepackage[latin9]{inputenc}
\usepackage{amsmath}
\usepackage{amssymb}
\usepackage{graphicx}

\makeatletter

\makeatletter

\usepackage{amsfonts}

\usepackage{times}

\usepackage{epstopdf}

\makeatletter

 \usepackage{epstopdf}
    \usepackage{times}
\makeatother

\makeatother

\usepackage{babel}
\begin{document}

\title{Effect of Interfacial Structural Phase Transitions on the Coupled
Motion of Grain Boundaries: A Molecular Dynamics Study}

\author{T. Frolov}

\maketitle
timfrol@berkeley.edu

Department of Materials Science and Engineering, University of California,
Berkeley, California 94720, USA
\begin{abstract}
In this work the coupled motion of two different phases of $\Sigma5(210)[001]$
grain boundaries were investigated by molecular dymamics simulations
of fcc Cu. The effect of interfacial structural phase transitions
is shown to have a profound effect on both the shear strength and
the nature of the coupled motion. Specifically,  the motion of the
two different phases is described by ideal coupling factors $\beta^{<100>}$
and $\beta^{<110>}$ that have different magnitudes and even signs.
Additionally, the shear strength for the two interfacial phases is
observed to differ by up to 40 \% at the lowest temperatures simulated.
The study demonstrates that grain boundary phases transitions may
have strong effects on the kinetics of microstructural evolution. 
\end{abstract}
Grain boundaries, coupled motion, grain boundary migration, grain
boundary structure

Grain boundary (GB) phase transitions are of profound fundamental
interest and can have a significant effect on macroscopic properties
of materials.\cite{Cantwell20141} Experimental studies suggest that
GB phase transitions play important role in abnormal grain growth
in ceramics,\cite{Dillon20076208} activated sintering\cite{Luo99}
and liquid metal embrittlement.\cite{Luo23092011} Despite recent
progress the effect of these transitions on the kinetics of microstructural
evolution is not well understood. 

Much of our knowledge about GBs comes from atomistic modeling. Atomistic
simulations have provided crucial insights into GB structure and mechanisms
of GB migration. Equilibrium and non-equilibrium GB properties like
free energies\cite{Broughton86a,Foiles2010,Foiles94,Frolov2012b}
and mobilities\cite{Schonfelder97,Karma2012,Upmanyu98,Janssens06,Trautt06a,Foiles06a}
were calculated as functions of misorientation and temperature.\cite{Balluffi95,Mishin2010,JJHoyt2013}
These properties are often assumed to change continuously with temperature
or chemical composition. However, direct studies of the effect of
discontinuous changes in GB structure at fixed misorientation on GB
migration have not been reported from atomistic simulations to date.

Recent molecular dynamic (MD) simulations reported first order structural
transitions in metallic grain boundaries.\cite{Olmsted2011,Frolov2013}
The discovery of these transitions were enabled by a new simulation
methodology. It was demonstrated that the transformations can be triggered
by point defects or temperature and gave rise to strong effects on
impurity segregation and diffusion.\cite{Frolov2013PRL} The system
studied in Ref. \cite{Frolov2013} provides a convenient model to
investigate the potential role of GB transitions on the kinetics of
GB coupled motion.

Coupled GB motion is characterized by a coupling factor $\beta$,
which is a ratio of the tangential and normal GB velocities. Molecular
dynamics studies of the coupled motion of symmetrical tilt $[100]$
GB in Cu demonstrated two modes of coupling, with ideal coupling factors
$\beta^{<100>}$ and $\beta^{<110>}$ given by\cite{Cahn06a}

\begin{equation}
\beta^{<100>}=2tan\left(\frac{\theta}{2}\right)\label{eq:beta100}
\end{equation}

\begin{equation}
\beta^{<110>}=-2tan\left(\frac{\pi}{4}-\frac{\theta}{2}\right)\label{eq:beta110}
\end{equation}
where $\theta$ is the misorientation angle. It was demonstrated that
GBs with $\theta$ closer to $0^{\circ}$ couple in the $<100>$ mode
, while GBs with $\theta$ closer to 90$^{\circ}$ degrees couple
in the $<110>$ mode. The discontinuous transition between the two
ideal branches of coupling was observed to occur around 36$^{\ensuremath{\circ}}$.
Dual coupling behavior was found for GBs near 36$^{\ensuremath{\circ}}$
degrees with the coupling factor spontaneously switching between the
two ideal modes during GB motion induced by an imposed constant shear-strain
rate in the MD simulations. However, the simulations used periodic
boundary conditions with the total number of atoms fixed, thus, prohibiting
possible structural transformations of the GBs. 

\begin{figure}
\includegraphics[height=0.35\textheight]{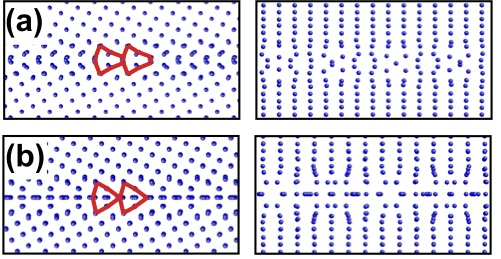}

\caption{(a) Split-kite and (b) filled-kite phases of $\Sigma5(210)$ grain
boundary calculated at 0K in Ref. \cite{Frolov2013} using an EAM
potential for Cu.\cite{Mishin01} \label{fig:GB_structures}}
\end{figure}

In this work we study coupled motion of different phases of $\Sigma5(210)[001]$
GBs in Cu.\cite{Mishin01} This is a typical high angle boundary with
misorientation angle of $53.13^{\circ}$. The two different atomic
structures for this boundary, identified in Ref. \cite{Frolov2013}
and referred to as ``split-kite\textquotedbl{} and ``filled-kite\textquotedbl{}
structures, are illustrated in Figure \ref{fig:GB_structures} (a)
and (b), respectively. The dimensions of the bicrystals with the two
GB phases were $5.1\times4.9\times16.5$ nm $^{3}$ , with the $[001]$
direction parallel $x$ , and GB planes normal to the $z$ direction.
Due to the difference in GB structures, the number of atoms in the
simulation blocks were 33677 and 33576 for split kites and filled
kites, respectively. Periodic boundary conditions were applied in
the $x$ and $y$ directions, while fixed boundary conditions were
used in the $z$ directions to implement shear deformation.\cite{Cahn06b}
Molecular dynamics simulations were performed using using the Large-scale
Atomic/Molecular Massively Parralel Simulator (LAMMPS) software package.\cite{Plimpton95}
MD simulations of coupled motion were performed in the temperature
range from 0.38-0.98 $T_{m}$, where $T_{m}$=1327 K is the melting
temperature for the Cu potential due to Mishin\cite{Mishin01} employed
in this work. To investigate the effects of the rate of the deformation,
the simulations of coupled motion were performed with shear rates
spanning an order of magnitude, with imposed shear velocities of 1m/s
and 0.1 m/s. 

\begin{figure}
\includegraphics[height=0.8\textheight]{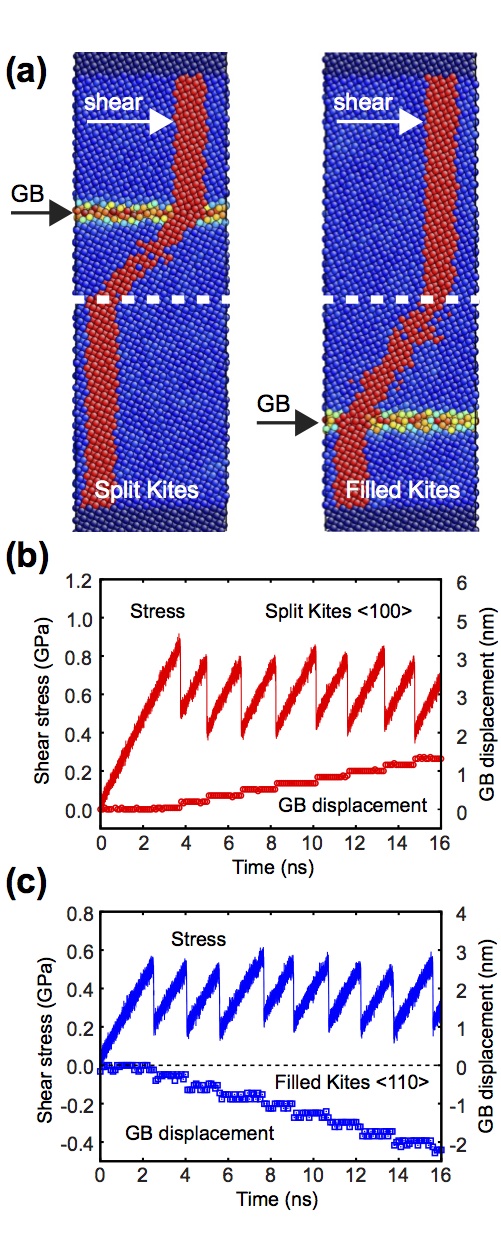}

\caption{(a) Bicrystals with split kites (left) and filled kites (right) GB
phases after 30 ns of coupled motion at 700 K with $v_{\parallel}=0.1$
m/s. Initially vertical, red stripe serves to illustrate the shear
deformation. Shear stress and grain boundary displacement during coupled
motion of split-kite (b) and filled-kite (c) phases. \label{fig:GB disp}}
\end{figure}

Figure \ref{fig:GB disp} (a) illustrates the bicrystals with split-kites
and filled-kite GB phases after 30 ns of the coupled motion. The initial
positions of the GBs are indicated by the white dashed lines. To track
the shape changes of the bicrystals, a column of atoms that were initially
vertical are colored in red. It is evident from the figure that the
part of the block swept by the GBs is sheared. While the upper grains
of both bicrystals moved to the right by the same amount as indicated
by the white arrows, the split kites GB (left) moved up and filled
kites (right) moved down. The amounts of the normal displacement observed
for the same simulation time are also different. 

Figures \ref{fig:GB disp} (b) and (c) show typical plots of the shear
stress and the GB displacements as functions of time at 700 K for
split kites and filled kites, respectively. Both phases move by characteristic
stick-slip dynamics, however the direction and magnitude of GB displacements
as well as the shear stresses are different for the two GB phases. 

\begin{figure}
\includegraphics[width=0.7\textwidth]{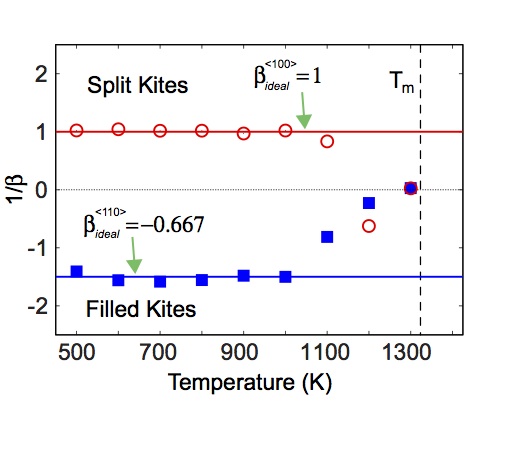}

\caption{Inverse coupling factor $1/\beta$ as function of temperature for
split kites (red circles) and filled kites (blue squares). Solid horizontal
lines correspond to ideal values $1/\beta^{<100>}$ and $1/\beta^{<110>}$
calculated using Eqs. (\ref{eq:beta100}) and (\ref{eq:beta110}).
The vertical dashed line indicates the bulk melting temperature $T_{m}=1327$
K. \label{fig:Coupling_factor}}
\end{figure}

From the knowledge of the imposed shear rate and the observed normal
displacements (Figures \ref{fig:GB disp} (b) and (c)) the value of
the GB coupling factor $\beta=v_{\parallel}/v_{\bot}$ was calculated
from MD simulations. Figure \ref{fig:Coupling_factor} shows the inverse
coupling factors $1/\beta$ for the two GB phases as functions of
temperature. The blue and red horizontal lines on the plot represent
values of the ideal coupling factors $\beta^{<100>}$ and $\beta^{<110>}$,
calculated using Eqs. (\ref{eq:beta100}) and (\ref{eq:beta110}).
It is evident from the figure that until about 0.83 $T_{m}$ the split
kite and filled kite phases move with ideal coupling factors, corresponding
to the $<100>$ and $<110>$ branches, respectively. For this range
of temperature and shear rates the GB structures are preserved during
the coupled motion and no spontaneous switching between modes of coupling
was observed. The absence of switching is different from results reported
in previous studies\cite{Cahn06b} indicating that the coupling behavior
is sensitive to the details of the atomistic GB structure. 

At temperatures above 0.83 $T_{m}$ coupling factors of both GBs become
non-ideal. The filled-kite phase shows increasing sliding until it
completely premelts at 1300 K, and the motion of the GB is purely
sliding in nature. The split-kite phase also becomes increasingly
disordered with its structural units disappearing around 1200 K. This
change in the GB structure correlates with the observation of a non-ideal
value and change in sign for the coupling factor  $\beta$. 

\begin{figure}
\includegraphics[width=0.7\textwidth]{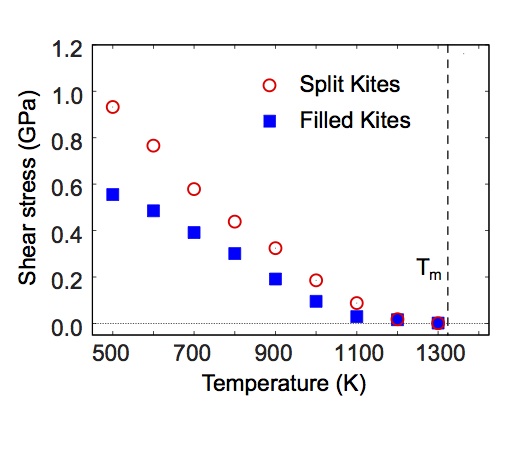}

\caption{Shear stress as a function of temperature for split kites (red circles)
and filled kites (blue squares) for $v_{\parallel}=1$ m/s. The vertical
dashed line indicates the bulk melting temperature $T_{m}=1327$ K.
\label{fig:Stress}}
\end{figure}

Finally, the average shear stress during coupled motion was calculated
for the two GB phases and is plotted as a function of temperature
in Figure \ref{fig:Stress}. While both stresses decrease with temperature,
the stress required to move filled kites is only about 60\% of that
for split kites. This shows that given the same driving force GB transitions
can noticeably increase or slow down GB motion.

The GB phases studied here differ significantly in atomic density,
and as a result the kinetics of these transitions are limited by diffusion
of atoms. It is thus quite possible that different GB phases may be
present simultaneously in equilibrium with each other or due to slow
kinetics. The mechanism of motion of such boundaries is currently
not clear, since different segments of the boundary will move in opposite
directions in the response to the same driving force. As a result,
such mixed-phase GBs can become less mobile with one of the phases
dragging the other. 

This study also motivates a systematic investigation of coupled motion
of different phases of other high angle GBs, to see weather the simultaneous
presence on different branches of coupling depending on GB structure
is a common phenomenon. Experimental measurements of coupling factor
for Al showed an abrupt transition from $<100>$ to $<110>$ modes
with increasing misorientation,\cite{Gorkaya09} confirming results
the previous MD studies for Cu.\cite{Cahn06b} The current investigation
shows that the switching of the GB coupling mode for the same grain
misorientation can be triggered by GB phase transitions. Thus, the
study suggests a possible indirect way to discover GB phase transitions
by measuring variations in the coupling factor at different temperatures
and impurity levels. 

In summary, coupled motion of split-kite and filled-kite phases of
a $\Sigma5(210)$ GB in Cu was investigated using atomistic simulations.
Both the coupling mode and GB shear strength were observed to depend
on GB phase for fixed misorientation angle. Specifically, the motion
of the phases is described by ideal coupling factors $\beta^{<100>}$
and $\beta^{<110>}$ that have different magnitudes and even signs.
Thus, the present simulation results demonstrate clearly that GB structural
phase transitions have a significant effect on kinetic properties
of GBs and in some cases may even reverse the direction of GB motion.\cite{Borovikov2013}

\section*{Acknowledgments}

The author is grateful to M. Asta for reading the manuscript and providing
valuable comments. The author also acknowledges support through a
postdoctoral fellowship from the Miller Institute for Basic Research
in Science at University of California, Berkeley.


\bigskip{}

\end{document}